                                \documentstyle[12pt,aasms4]{article}
                                \begin{document}     
                                \title{              
Seeing Galaxies Through Thick and Thin: II. Direct Measures of Extinction in 
Spiral Disks Through Spectroscopy of Overlapping Galaxies
				        }                    
                                \author{             
Donovan L. Domingue, William C. Keel\altaffilmark{1} \& 
Raymond E. White III\altaffilmark{1} 
                                }
                    
                                 \affil{
Department of Physics and Astronomy,     
University of Alabama,               
Box 870324, Tuscaloosa, AL 35487-0324;
Electronic mail: domingue@hera.astr.ua.edu, keel@bildad.astr.ua.edu, 
white@merkin.astr.ua.edu   
                                }
\altaffiltext{1}{Visiting astronomer, Kitt Peak National Observatory, National
Optical Astronomy Observatories, Operated by AURA, Inc. under cooperative 
agreement with the NSF.} 
                                \begin{abstract}
We use slit spectroscopy of overlapping pairs of galaxies to directly determine
the extinction in disks of foreground spiral galaxies. The Doppler shifts of
pair members are determined via cross-correlation and their relative correlation
amplitudes are used to separate their contributions to the combined spectra in 
regions of overlap. This spectroscopic approach is less subject to stringent 
symmetry constraints than our previous purely photometric analyses. Extinctions
of foreground members were obtained for 6 of the candidates in our sample of 18
mostly spiral/spiral pairs, when the signal to noise and velocity difference
were suitable. In agreement with our previous imaging results, we find that the 
extinction in interarm regions is very modest, typically $A_B\sim0.1$ mag 
(corrected to face on), while spiral arms exhibit higher extinctions of 
$\sim0.3$ mag.
                                  
				\end{abstract}
                                                
				\clearpage
                                \section{
Introduction                             
                                 } 
Determining the opacity of spiral galaxy disks is essential to
understanding how the dust content in disks can affect luminosity estimates, 
star formation rates, and the ability to see through these disks when 
observing distant objects such as QSOs. 
The inclination--surface-brightness test is one of the oldest methods 
(Holmberg 1958) used to determine whether spiral galaxies are largely 
transparent or opaque:  opaque disks have surface brightnesses independent 
of inclination, while transparent disks are brightest when seen edge-on.
A more refined version of this test used by Valentijn (1990)
led to the claim that spiral galaxies are opaque. This result seems contrary to 
examples of objects seen through spiral disks and the fact that we are able to 
observe galaxies by looking beyond the Milky Way. Reassessments of Valentijn's 
(1990) work by Burstein et al. (1995) and Huizinga (1994) have shown that 
sample selection effects can mask any relationship between inclination and
surface brightness. 

White \& Keel (1992) initiated a campaign to determine $directly$ 
(rather than statistically) the opacity
of spiral disks by imaging partially overlapping galaxies.
The non-overlapping parts of the galaxies in such systems
are used to estimate how much light from the background galaxies is absorbed
in passing through the foreground galaxies (White \& Keel 1992; 
Keel \& White 1995; White, Keel \& Conselice 1996, 1999).
The success of this differential photometric technique is strongly dependent 
on the symmetry of each
of the overlapping galaxies, since the errors in the deduced optical depths are
dominated by systematic errors due to deviations from structural symmetry.

In our photometric analysis, 
the symmetry requirements for both background and foreground components of a 
pair limits the number of useful pairs from our observed sample (White, Keel \& 
Conselice 1999). Taking spectra of the overlap regions allows the symmetry 
requirements for the foreground galaxy to be relaxed, since we can directly 
identify photon ownership, due to the velocity difference between the two 
galaxies. A Fourier cross-correlation of the spectra shows peaks at the 
velocities corresponding to the redshifts of each of the pair members. The 
relative Fourier amplitudes of the two peaks should be equivalent to the relative 
intensities of the foreground and background galaxies in the overlapping region. 
This velocity decomposition, combined with photometry of a symmetric background 
galaxy region, gives us an estimate of the foreground disk opacity. Here we 
present new opacity measurements using this technique for arm and interarm regions 
in six spiral galaxies. The relaxed symmetry requirements for the foreground 
galaxy allow us to analyze spirals which are not grand design.

                                \section{
Method Formalism               
                                }
We first describe the differential photometric technique employed in our 
previous imaging studies, in order to contrast it with the present 
spectroscopic study.
A partially overlapping pair of galaxies is illustrated in Figure 1.
In the region of overlap, the surface brightness of the foreground galaxy 
is $F$ and the unattenuated surface brightness of the background galaxy is $B$. 
Light from the background galaxy is attenuated while passing through the
foreground galaxy, so what is actually observed in the overlap region is 
$\langle F+Be^{-\tau}\rangle$, where $\tau$ is the optical depth in the 
foreground galaxy and the angle brackets are used to emphasize that
the enclosed quantity is the basic observable.
We use symmetric regions of both the foreground and background galaxies to
$estimate$ $F$ and $B$ in the overlap region, since we cannot directly 
decompose them with imaging; 
the estimates of $F$ and $B$ are denoted $F^\prime$ and $B^\prime$.
We then use purely differential photometry to derive an estimate of $\tau$,
denoted $\tau^\prime$:
\begin{equation}
e^{-\tau^\prime} = { {\langle F+Be^{-\tau}\rangle - F^\prime} 
\over B^\prime }.
\end{equation}
That is, the estimate of the foreground light in the overlap region is first
 subtracted from the total
light in the overlap region; the result is then divided by the estimate of 
unattenuated background light to derive an estimate of $e^{-\tau}$.

In contrast, our spectroscopic technique is illustrated in Figure 2, where a 
slit is seen crossing the overlap region and intersecting both galactic centers.
In this case, if the radial velocity difference between the two galaxies is
large enough, the foreground light $F$ and the attenuated background light
$\langle Be^{-\tau}\rangle$ $can$ be spectroscopically distinguished in the 
overlap region, in contrast to the purely photometric case above.
In this case, all we need is an estimate $B^\prime$ of the $un$attenuated  
background galaxy's surface brightness $B$, derived from a symmetric region of
the background galaxy.
Our estimate of $\tau$ in the overlap region can then be deduced with a
modified version of equation (1):
\begin{equation}
e^{-\tau^\prime}= {\langle Be^{-\tau}\rangle \over B^\prime },
\end{equation}
since $F$ and $\langle Be^{-\tau}\rangle$ are spectroscopically separable. We 
require only global symmetry in this case since $B$ and $B^\prime$ may be 
averaged over substantial areas along the slit.

Puxley and James (1993) also exploited velocity separation of galaxy pairs in
this context, using emission lines from HII regions. Our use of stellar 
$absorption$ lines offers important advantages. The background continuum is 
smoothly distributed across large regions. Thus, we are not subject to point to
point variations in opacity and are not biased against high opacity regions.

The cross-correlation method as applied will use all absorption features
in our wavelength range (3600-5200 \AA) of the template and galaxy spectra. This
has the advantage of using lines blended by galaxy velocity dispersions in addition to easily identifiable lines. When they are taken into 
account, those blended lines may contribute the most signal to the
cross-correlation. The combined spectra of two galaxies in the overlap region 
is compared to our templates. The absorption features of the individual galaxies
contribute to the correlation with differing strength due to their intensity
levels. The brightest spectrum of the two will dominate and be most easily 
correlated, thereby producing the largest peak at its respective redshift.

                                \section{
Observations and Data Analysis  }

     To explore how well Doppler separation of the light from the foreground
and background galaxies would allow us to measure the extinction in
overlapping pairs, we obtained optical spectra of 18
such pairs (see Table 1) using the 2.1m telescope and GoldCam spectrometer at
Kitt Peak National Observatory in 1995 November. We concentrated on the 
line-rich spectral region from 3600-5200 \AA, with the full spectral range
falling on the $3072 \times 512$-pixel useful region of the Ford CCD, 
set comfortably wider than this (3100-6800 \AA) so that defocusing at the ends
of each spectrum would not compromise our cross-correlation results.
The slit width of 150$\mu$ (1\farcs9) gave a resolution of 4.1 \AA\ (FWHM)
with pixels subtending 0\farcs78 by 1.25 \AA. A slit length of 5\farcm0 gave 
clear sky around the pairs to permit sky-subtraction for all members of the 
sample but NGC 4567/8. Most objects received total exposures of 120 minutes, 
intended to give usable count rates within at least the overlap region as a 
whole and, in several cases, for several independent bins within this region. 
Each object exposure was immediately followed by one of a
HeNeAr comparison lamp, so that an individual wavelength solution could be 
generated for each observation. The rms deviation of comparison-line wavelengths
about the adopted solution was typically 0.14 \AA, so that the contribution of
the calibration step to radial-velocity errors would be typically less than 
10 km s$^{-1}$. 
Vignetting along the slit was measured and removed through an observation
of the twilight-sky zenith.
The velocity zero point was confirmed through nightly observations of
the IAU radial-velocity standard stars 
BD $+28^\circ$ 3402  (F7 V, -36.6 km s$^{-1}$),
HD 213947  (K4 III, +16.7 km s$^{-1}$), or
HD 90861 (K2 III, +36.3 km s$^{-1}$).
We also observed M32 at the start of the run, as an additional cross-correlation
reference.

Supporting $B$ and $I$ images for most of these pairs are available from
the CTIO 1.5m, KPNO 2.1m, or Lowell 1.1m telescopes (as described by
White, Keel, \& Conselice 1998). For two systems in which the slit
position did not cross the nuclei, we obtained an additional brief
spectrum of the nuclei, to verify the systemic radial velocity of each
galaxy.

Each sky-subtracted spectrum in our sample was cleaned using IRAF to get rid of
cosmic rays and remove columns where sky lines did not subtract well. Removal 
of pixel columns at wavelengths greater than 6500 \AA\ was done to remove poor 
data. Since the line-rich region contributing most of the cross-correlation 
power is below 5200 \AA, no interesting information was lost.
Averages were taken in the spatial direction across galaxy nuclei and turned 
into one dimensional spectra. The 1-D spectra were cross-correlated, using the
IRAF task FXCOR, with the radial velocity standard spectra to get redshift
estimates and determine which standards worked best for each galaxy. The spectra
of M32 and HD213947 were determined to be the most useful references but 
BD $+28^\circ$ 3402 was also used as a check on velocity estimates. After the
galaxies with previously known redshifts were shown to agree with our estimates,
heliocentric redshifts were obtained for the galaxies with unpublished 
radial velocities.

To analyze the area of combined spectra, the overlap region was located in 
angular distance from the galaxy nuclei along the position angle of the slit.
Extracted spectra were usually the sum of four detector pixels (3\farcs1)
along the slit, to increase the signal to noise.
Whenever possible, adjacent regions overlapped each other by two 
pixels. These 1-D spectra were cross-correlated with the radial velocity 
standards and double peaked profiles were sought near the 
respective velocities of the pair members. When a correlation proved 
unsuccessful, the sizes of the 1-D spectra were reevaluated and summed over a 
smaller region until results were achieved or options were exhausted. Final 
region selections are shown in Figures 5 through 11.

If observed cross-correlation peaks were located at the expected redshifts, 
their relative amplitudes were compared. Averages of relative amplitudes from 
correlations with two different standards were calculated when possible.
The relative amplitudes were taken as the fraction of foreground to background 
light. Association of the peaks with the position of galaxies was determined 
by matching velocities, since foreground and background contributions switched 
dominance as a function of distance from the respective nuclei. In one pair, 
NGC 7268, the foreground galaxy has a higher recession velocity
than that of the background galaxy by $\sim400$ km s$^{-1}$. It is 
interesting to note that in our first clean photometric case, AM1316-241, the
recession velocity of the foreground galaxy exceeded that of the background
galaxy by an even greater amount, $\sim700$ km s$^{-1}$ (White \& Keel
1992).

Once the fractional foreground contribution was determined, two methods were 
used to estimate the fluxes from symmetric areas of the background galaxy.
First, when the slit encompassed the necessary symmetric background region, 
the galaxies' spectra were averaged over all wavelengths to create an 
intensity profile of the pair. The intensities in the symmetric background
regions in the slit were then estimated to produce opacity measures as 
described above. 
Second, we used images of the pairs to estimate the 
unabsorbed contribution of the background galaxy from symmetric, 
nonoverlapping regions. 
This method is a hybrid of our purely photometric and our purely spectroscopic 
techniques.  When we use both methods, we refer to the purely spectroscopic
technique as the spectral method and the hybrid as the spectrum/image
method.

The velocity per pixel resolution of our spectra proved to be a limiting factor
in obtaining double velocity peaks in the cross-correlations. Since 
most of the galaxies differed in velocity by less than $\sim$ 300 
km s$^{-1}$, the peak with lower correlation amplitude could be located within 
the span of the higher amplitude peak. A higher spectral dispersion would 
produce greater separation in the peaks. This blending was usually the reason 
why a candidate pair failed to yield opacity measures. In some cases the orientation of the galaxy disks produced an advantageous increase in velocity separation at the regions of interest which allowed measures to be made for
NGC 1738/9 despite the pair's relative velocity. Other failures were 
due to obtrusive emission lines in the desired region of the spectra or to low
signal to noise. The role of signal to noise was investigated with synthetic 
composite spectra having varying S/N ratios. Decreasing the signal to noise
in a spectra with two equal correlation amplitudes at a separation of 500 km 
s$^{-1}$ made distinguishing peaks more difficult while the lowest ratios caused
FXCOR to be unable to find any peaks at the proper locations.

Metallicity effects among the galaxy pair members were found to be unimportant
since the depth of absorption lines is not as important as their placement with
respect to the template spectrum. Our templates from 
the spectra HD213947, a K star, and M32 with a K star dominated spectrum show
metallicity differences by way of line depths. Both of these templates were
used in the cross-correlations and the very similar results were averaged and
reported. An increased relative line depth would improve the correlation if it were large enough to stand out in a noise filled spectrum, so our signal to
noise ratio is a larger concern to achieve results. The two galaxies of largest
morphological difference in our useful sample are the members of the NGC 7268 
pair. Using FXCOR to cross-correlate non-overlapping portions of this pair with
one template reveals the same level of correlation for an SBbc and an S0, therefore metallicity effects are ignored. Since we also observe the 
non-overlapping portions of the galaxies within the spectroscopic slit, we are 
only sensitive to possible asymmetries in line strengths.
 
We used synthetic composite spectra to calibrate the mapping between 
ratios of Fourier amplitudes to foreground/background intensity ratios. 
We found that the mapping of this ratio from real space to Fourier space 
is dependent on the velocity resolution of the input spectra. 
Artificial spectra with different redshifts were created with MK1DSPEC 
and combined with known relative contributions of background and foreground light. 
The resulting cross-correlation amplitudes were then compared to the input 
intensity ratios.  
Artificial spectra were created with a velocity resolution comparable
to that of our real data, which have 75 km s$^{-1}$ pixel$^{-1}$. The test 
Fourier amplitude ratios of the two peaks were found to be within 10\% of the 
input intensity ratios, independent of the value of the ratio. Figure 4 plots 
the input continuum ratio against that inferred from FXCOR for the 75 km
s$^{-1}$ pixel$^{-1}$ tests with various velocity space separations. Nearly 
all observed amplitude ratios are between 1 and 2, with only two regions greater
than 3, so the expected errors from the Fourier measure should all fall near 
5\%. Similar tests were performed by shifting the spectrum of M32 to two
different velocities and combining them into one spectrum. Those results show
that correlation amplitude blending for separations of less than 500 km s$^{-1}$ causes ratios to be less accurate as is also shown in Figure 4. With this
analysis, pairs still fall within an acceptable 10\% error. These effects are smaller than our measurement errors so we do not apply corrections.  
                                 \section{
Results                               }

Individual galaxy pairs which gave useful opacity measurements are discussed below.
The extinctions are reported in magnitudes, where the relation between magnitudes 
of extinction $A$ and optical depth $\tau$ is $A$=1.086$\tau$. Our measured values 
correspond roughly to $A_B$, since most of the strong absorption features giving 
cross-correlation signal lie in the $B$ band. 
Other pairs in our sample lack discernible double peaks in the 
collected spectral cross-correlations. Table 2 contains a summary of properties 
of the seven pairs with useful results followed by those galaxies which did not
yield results; the redshifts listed are those calculated with
FXCOR unless otherwise noted. All morphological types, $R/R_{25}^B$ values, and 
noted redshifts were obtained from the NASA/IPAC Extragalactic Database (NED), the 
ESO-LV catalog (Lauberts \& Valentijn 1989), or the RC3 (de~Vaucouleurs et al. 1991). 
Table 3 lists the face-on corrected extinctions, along with corresponding regions 
and averages. Negative extinction values in the table indicate the level of systematic 
error due to departures from symmetry in the associated background galaxies.

					\subsection{
ESO 054642-2534.4 			}

ESO 054642-2534.4 is comprised of two SBs in the cluster AM 0546-253 (see
Figure 5) and is our most easily distinguished 
spectral pair, due their approximate velocity separation of 1300 km s$^{-1}$. 
Results were obtained for four regions with a two pixel wide overlap,
corresponding to 1\farcs56 of shared width. These regions span nearly 2.5 
kpc, using the galaxy redshift distance ($H_0=75$ km s$^{-1}$ Mpc$^{-1}$). 
The four areas sweep across a background arm/ring and a foreground arm/ring. 
The spectral intensity profile method yields an average extinction of 
$A_B\approx0.25$, ranging from 0.08 to 0.48; 
the $B$ image/spectrum method yields an average $A_B\approx0.11$. 
Face-on corrected values are 0.17 and 0.07 for the two methods, respectively,
after dividing by the axial ratio $a/b=1.47$. Interarm material in the
foreground galaxy is revealed to have average extinctions of $A_B=0.2$ and 0.15
through inspection of three of the regions using the two methods defined above.
	
                       \subsection{
ESO 064906-3517.3			}

In ESO 064906-3517.3, an Sa-Sab pair, we found three regions which give double peaked 
cross-correlations (see Figure 6). These regions are four pixels (3\farcs1)
wide and span an area including some foreground spiral arm. 
The foreground and background galaxies contribute comparable amounts of light
in each region. Extinctions average to $A_B=0.18$ using the spectrum intensity
profile method. These data imply a face-on value of $A_B=0.11$, assuming a
simple cosine inclination dependence and the observed foreground galaxy axial
ratio $a/b=1.61$. Face-on values for the regions identified as being in a
foreground arm average to be $A_B=0.15$.

					\subsection{
MCG -02-58-011			}

MCG -02-58-011 is comprised of an edge-on foreground SBc, oriented E-W,
projected against a highly inclined blue late-type spiral, oriented N-S. 
Reddened knots are seen in the disk oriented E-W only in the overlap
region; these knots are otherwise similar to those in the disk oriented N-S, 
which leads to the conclusion that the edge-on 
E-W SBc is in the foreground. We found six regions with discernible
light contributions from both pair members (see Figure 7); 
these regions are four pixels wide and each shares two pixels. 
Unfortunately, these regions have no unobscured symmetric counterparts, so  
we cannot accurately estimate opacities. 
We can merely state that the foreground member is not completely opaque.

					\subsection{
NGC 1738/9				}

NGC 1738/9 comprises an SBbc-Sbc pair for which our single spectroscopic slit position
did not pass through background areas symmetric to our regions of interest
(see Figure 8), due to geometric constraints.
We were however able to estimate opacities despite the velocity separation being 
less than 100 km s$^{-1}$. 
Cross-correlation of spectra from the overlap region proved to be easiest with 
the nuclear spectrum of NGC 1738. In combination with the $B$ image/spectrum 
method, three one pixel (0.2 kpc) wide regions reveal an average apparent 
$A_B=0.76$ for these arm regions. 
Dividing by the axial ratio of the foreground galaxy NGC 1739 ($a/b=1.95$)
yields a face-on corrected value of $A_B=0.39$, which is in general agreement
with our previous photometric analysis of this pair (White, Keel, \& Conselice 1998). 
 
					\subsection{
NGC 3088				}

NGC 3088 consists of a nearly face-on foreground S0 projected against a background, 
nearly edge-on spiral galaxy.  We found four useful regions that are four pixels 
(3\farcs1) wide and share a width of two pixels with their neighbors (see Figure 9). 
Expected asymmetries in an edge-on spiral complicate measurements for
this pair. Negative values indicate that some obscured regions behind the S0 
have a greater brightness than geometrically symmetric regions in the
background member. Although the negative values are unphysical, they are
presented here and in Figures 12-14 for completeness. The S0 has average disk extinctions of $A_B=0.15$ using the 
spectrum method and $A_B=-0.30$ using the spectrum/image method. 
Face-on-corrected values are 0.13 and -0.27, respectively, given an axial
ratio of 1.12.
 
					\subsection{
NGC 6365				} 

The NGC 6365 pair consists of a nearly edge-on Sdm (NGC 6365B) projected against
a SBcd galaxy (see Figure 10).  We analyzed five regions, each one pixel wide,  
lying along NGC 6365B. The apparent extinctions derived from this spectrum
method indicate that NGC 6365B is the foreground galaxy which agrees with
the respective redshifts, however the decision is complicated by a region on 
NGC 6365B that appears diminished in brightness by a possible arm of the face-on
member. In contrast there appears to be an arm of NGC6365A peeking from
behind the edge-on member. The interpretation, that NGC 6365B is the foreground, is favored for meaningful extinctions. Because of spectral similarities, a cross-correlation with the NGC 1739 spectrum was most useful. 
The foreground galaxy has average face-on corrected interarm extinctions of $A_B=0.15$ and 0.17, for the spectrum and spectrum/image methods, respectively
(and an axial ratio of $a/b=3.70$).

					\subsection{
NGC 7268				}

The galaxy pair associated with NGC 7268 (see Figure 11)
has a foreground SB ringed galaxy 
with a recession velocity 400 km s$^{-1}$ greater than that of the 
background S0/E galaxy. Results were obtained for two regions,
each 1\farcs56 ($\sim1$ kpc) wide and not sharing any pixels. 
The location of these areas with respect to galaxy features appears to 
be in the outer ring of the foreground barred galaxy. 
The average ring extinctions are $A_B=0.53$ (using the intensity profile method)
and $A_B=0.30$ (using the $B$ image/spectrum method). 
The face-on corrected values are $A_B=0.35$ and $A_B=0.21$, given an
axial ratio of $a/b=1.50$.
 
                                \section{
Summary and Discussion
                                }

We have used slit spectroscopy of overlapping pairs of galaxies to 
determine directly the extinction in foreground spiral disks.
Our spectroscopic technique allows us to relax the symmetry requirements
of our previous imaging studies. In agreement with our previous imaging results,
we find that the extinction in interarm regions is typically $\sim0.1$ mag in 
blue (corrected to face on), while spiral arms exhibit extinctions of $\sim0.3$
mag. In this sample, rings exhibit the same level of extinction as spiral arms.
These numbers are calculated as an average of both methods used in this paper. 
Face-on extinctions are estimated by dividing by axial ratios and therefore can
be overestimates since realistic cases have finite thickness and clumped 
absorbers. The values obtained can be expected to have errors of at least 5\%
due to cross-correlation uncertainties but values will be most affected by
unwanted asymmetry in the background light comparisons.

A plot of uncorrected extinctions as a function of axial ratio (Fig. 12), for 
data obtained in this paper and from White, Keel, \& Conselice (1999), reveals a
great deal of scatter amongst arm regions. This comparison reveals no
inclination dependence in the apparent extinctions.  The separation of arm
and interarm values is clear. A similar comparison of face-on extinctions with 
galaxy type (Fig. 13) also reveals scattering of values for spiral arms and a 
more consistent measurement for the interarm regions. The individual internal 
structures of galaxies may play an important role in the scatter. In Figure 14,
extinctions as a function of $R/R_{25}^B$ are plotted and show a suggestion of
decreasing values with increasing radius for interarm regions while arm region
extinctions show no dependence on radius. The outlying points at the lower left 
are from the S0 galaxy, NGC 3088 which has a possibly asymmetric edge-on galaxy
in the background.

This spectral decomposition technique is a viable way of determining disk 
opacities. As discussed by White, Keel, \& Conselice (1999), these results 
suggest that galaxy extinction is not the cause of the QSO ``cutoff" and that 
intrinsic galactic luminosities are not as underestimated as they would be if 
spirals were completely opaque.  

This research has made use of the NASA/IPAC Extragalactic Database (NED) which
is operated by the Jet Propulsion Laboratory, California Institute of
Technology, under contract with the National Aeronautics and Space 
Administration. This research has also made use of {\sl SKYVIEW}. D.D. thanks 
Guy Purcell and Victor Andersen for helpful technical support. We thank Paul 
Eskridge for observing NGC 3088 and NGC 6365 with the 1.3 meter MDM. 

                                  
                                \clearpage
                                \title{       
Figure Captions                               
                                }


                               \figcaption{
Schematic of differential photometric technique for estimating extinction
in a foreground spiral of a partially overlapping pair of galaxies.
				}
                               \figcaption{
Schematic of spectroscopic technique for estimating extinction
in a foreground spiral of a partially overlapping pair of galaxies.
				}

                               \figcaption{
Sample cross-correlation results from FXCOR can be seen to exhibit two separate
peaks near the Gaussian fit. They are located at the respective redshifts of 
the galaxies and the peak heights represents the intensity contributions of the
pair members.
				}

                               \figcaption{
Plot of estimated accuracy for foreground/background light contributions 
obtained from FXCOR for synthetic spectra and M32 spectra with $\sim75$ km s$^{-1}$ pixel$^{-1}$ velocity resolution. Various velocity separations are
plotted.  
				}


                               \figcaption{
$B$ band image of ESO 054642-2534.4 with placement of spectroscopic slit 
illustrated along with the four numbered areas in Table 3. The regions are 
approaching a resonance ring but are primarily composed of interarm material.
Regions are not shown to overlap for purposes of clarity.
				}

                               \figcaption{
Digitized Sky Survey image of ESO 064906-3517.3 with placement of spectroscopic 
slit illustrated along with the three regions crossing a spiral arm numbered as
in Table 3. Regions are not shown to overlap for purposes of clarity.}

                               \figcaption{
$B$ band image of MCG -02-58-011 with placement of spectroscopic slit 
illustrated along with the six numbered areas in Table 3. The regions surround 
the nucleus of the vertical galaxy which is possibly the background member. 
They are located such that symmetric counterparts are represented by the
regions opposite the nucleus and prevent reasonable estimates of the extinction.
Regions are not shown to overlap for purposes of clarity.
                                 }

                               \figcaption{
$B$ band image of NGC 1738/9 with placement of spectroscopic slit illustrated 
along with the three numbered areas in Table 3. These regions are in a spiral 
arm of NGC 1739 which can be seen cutting out the light of NGC 1738.  
				}

                               \figcaption{
$B$ band image of NGC 3088 with placement of spectroscopic slit 
illustrated along with the four numbered areas in Table 3. The regions are near 
the edge of the S0. Regions are not shown to overlap for purposes of clarity.
 				}

                               \figcaption{
$B$ band image of NGC 6365 with placement of spectroscopic slit 
illustrated along with the five numbered areas in Table 3. The regions surround
the disk of the inclined galaxy.
				}

                               \figcaption{
$B$ band image of NGC 7268  with placement of spectroscopic slit illustrated 
along with the two numbered areas in Table 3. The regions are in a dusty
resonance ring of the foreground galaxy.  Regions are not shown to overlap for 
purposes of clarity.
				}

                               \figcaption{
Plot of apparent extinction as a function of axial ratio (a/b) for the ensemble
of galaxies reported on as well as those from White, Keel, \& Conselice (1999).
Filled circles represent the sum of arm regions for each galaxy and open
circles represent the sum of interarm regions. NGC 3088 is represented by 
plus signs in all figures.
				}
				
                               \figcaption{
Plot of face-on-corrected extinction as a function of galaxy type for the 
ensemble of galaxies reported on as well as those from White, Keel, \& 
Conselice (1999). Filled circles represent the sum of arm regions for each
galaxy and open circles represent the sum of interarm regions.
				}

                               \figcaption{
Plot of face-on-corrected extinction as a function of the ratio of radius to
de~Vaucouleurs radius for the ensemble of observed galaxy regions
reported on as well as those from White, Keel, \& Conselice (1999). Filled
circles represent arm regions and open circles represent the interarm regions.
The plus signs represent the regions from the NGC 3088 pair.
                                      }

                                
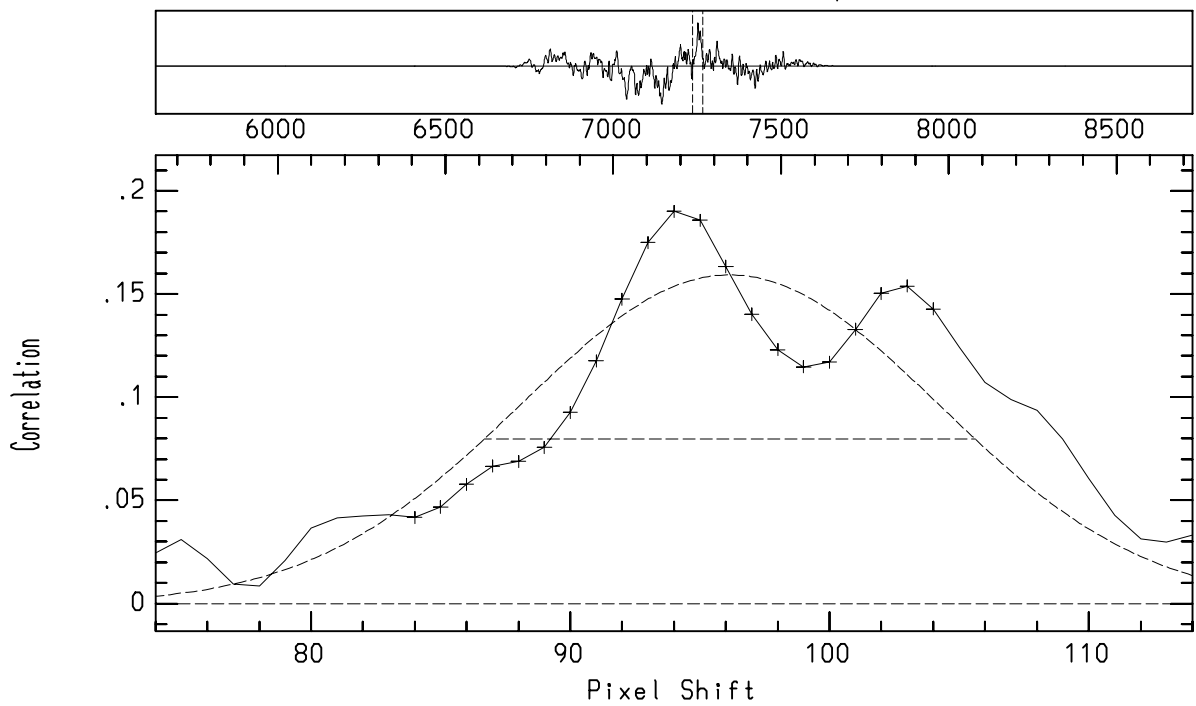
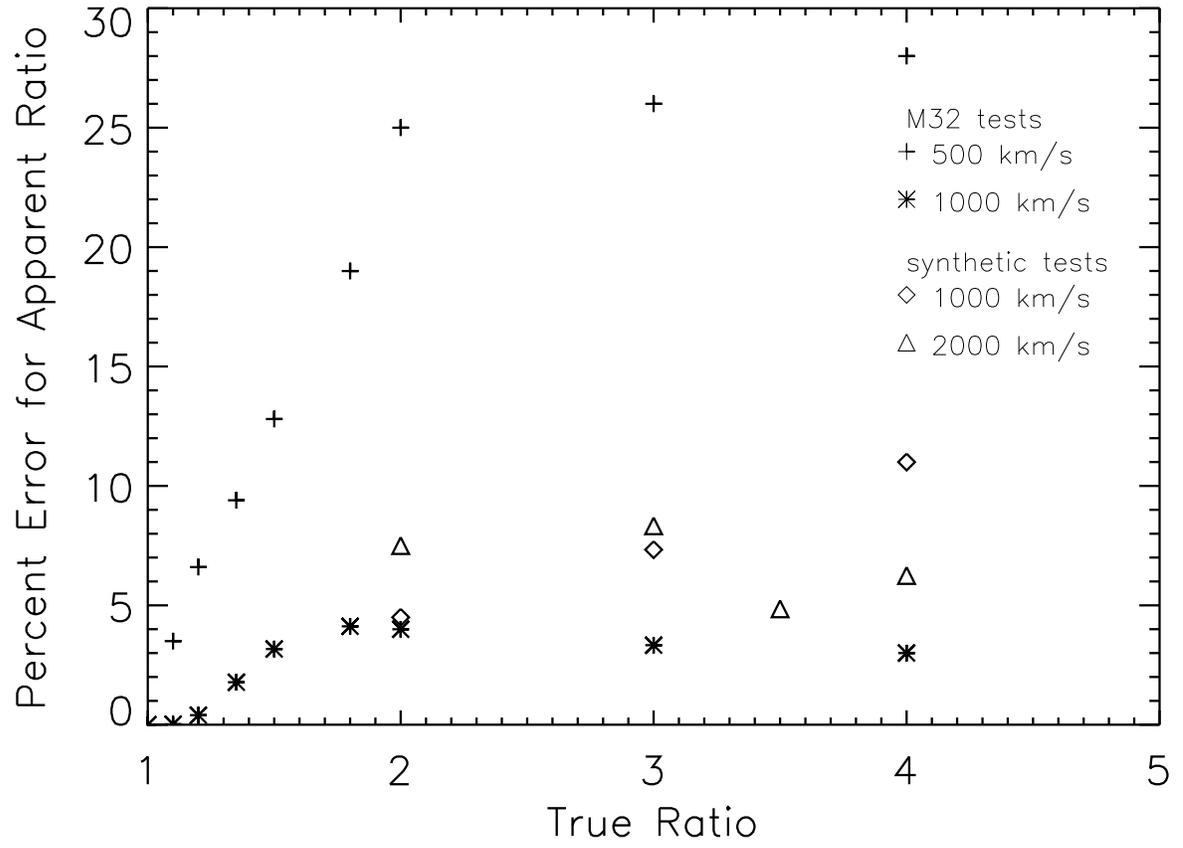
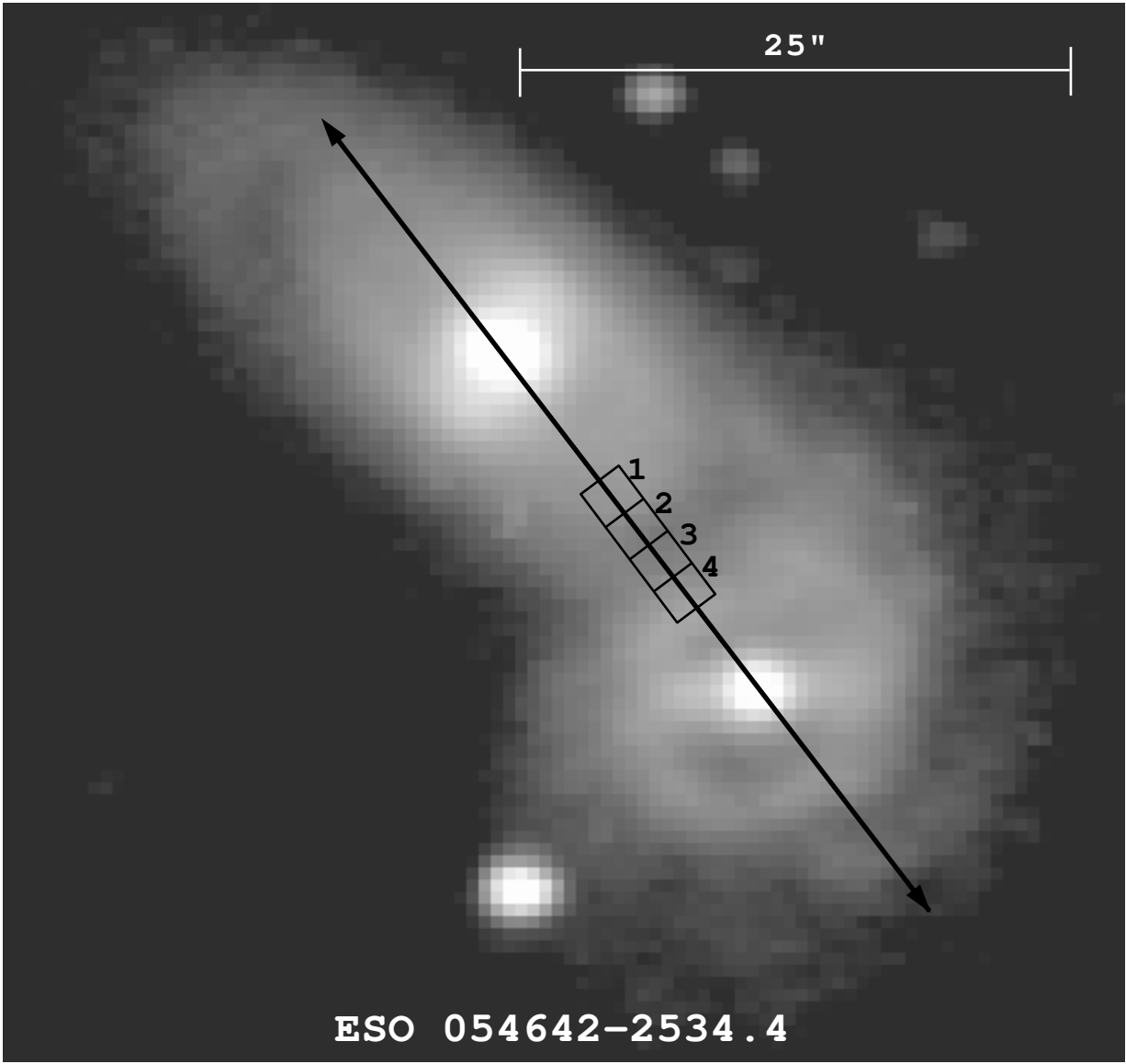
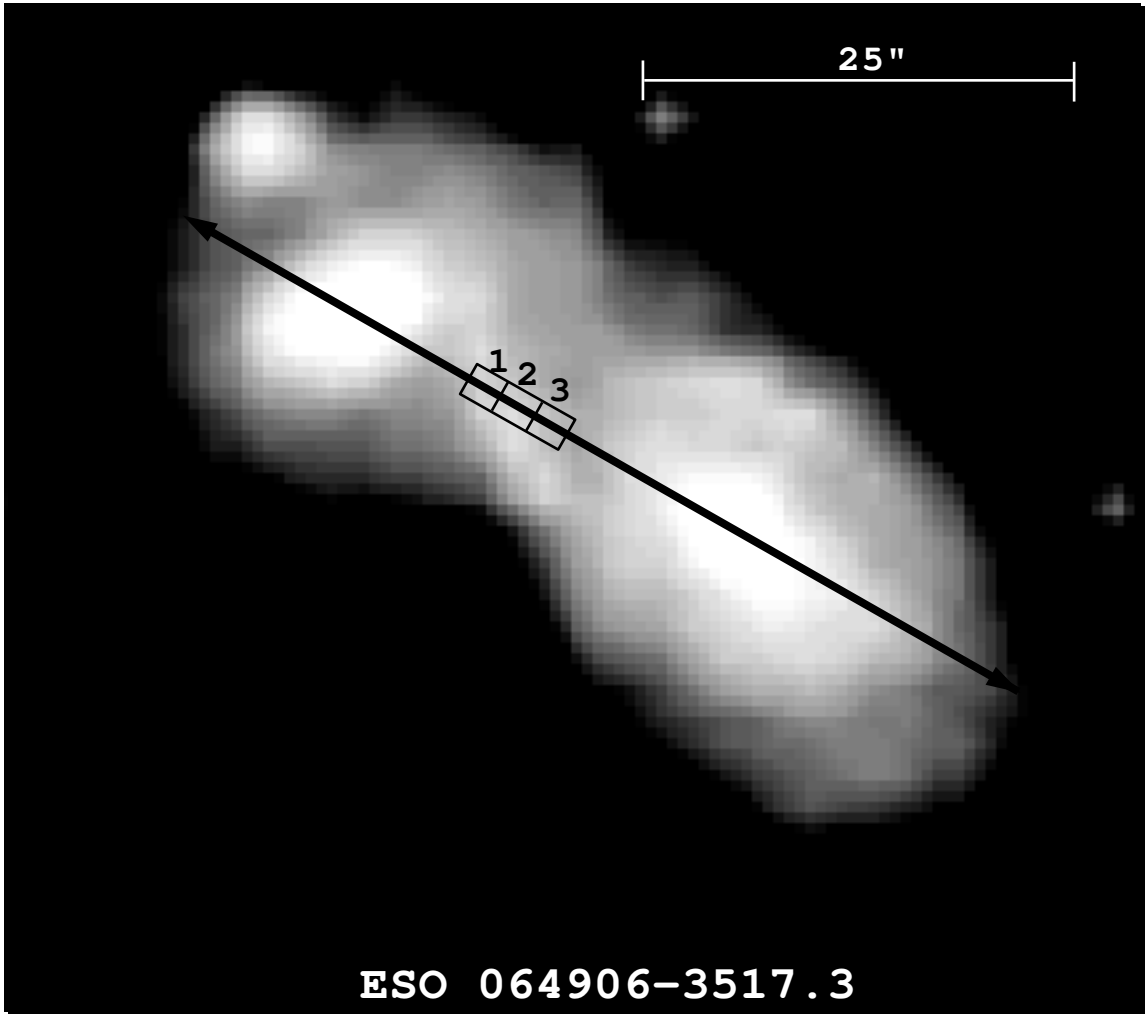
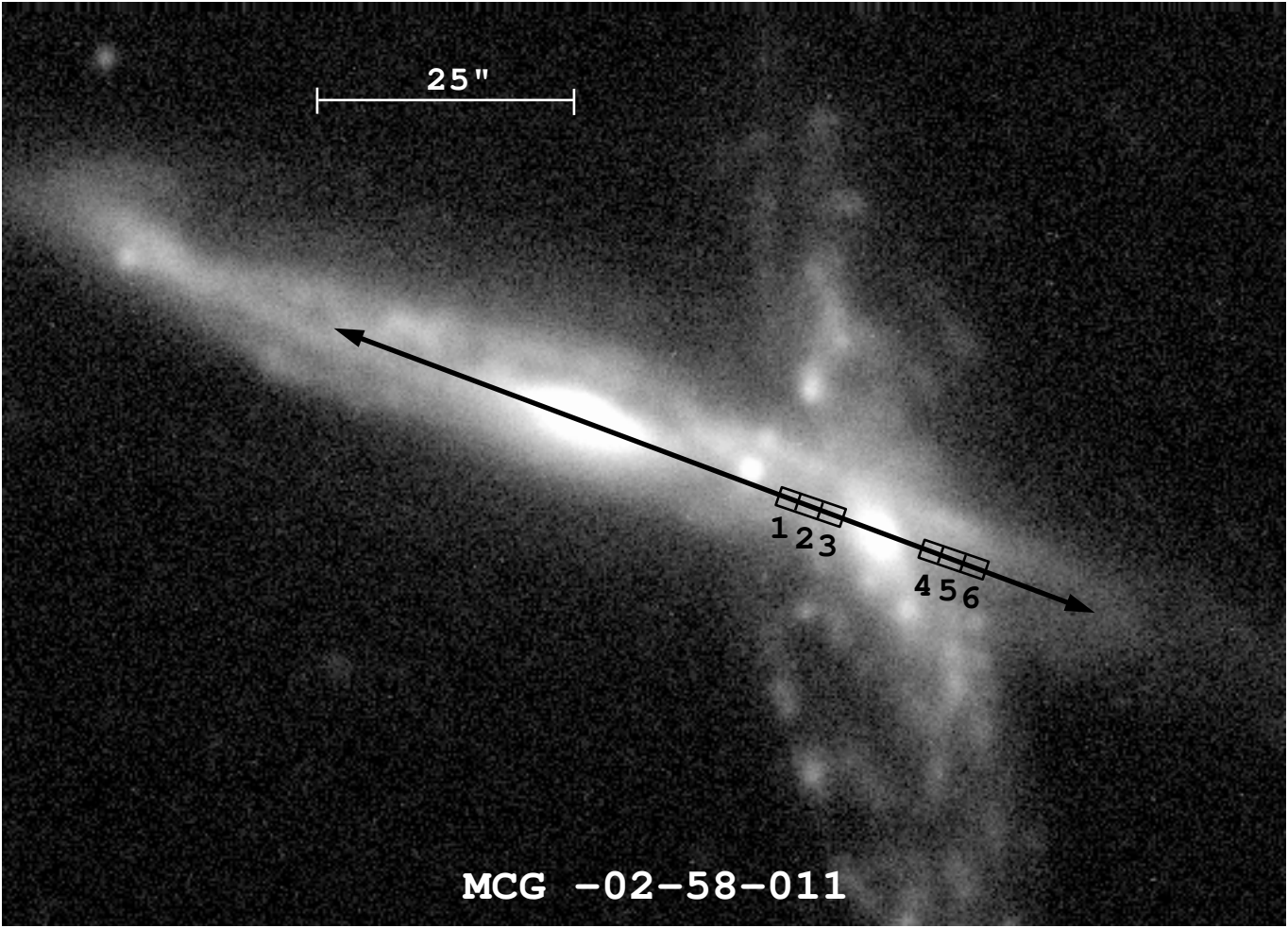
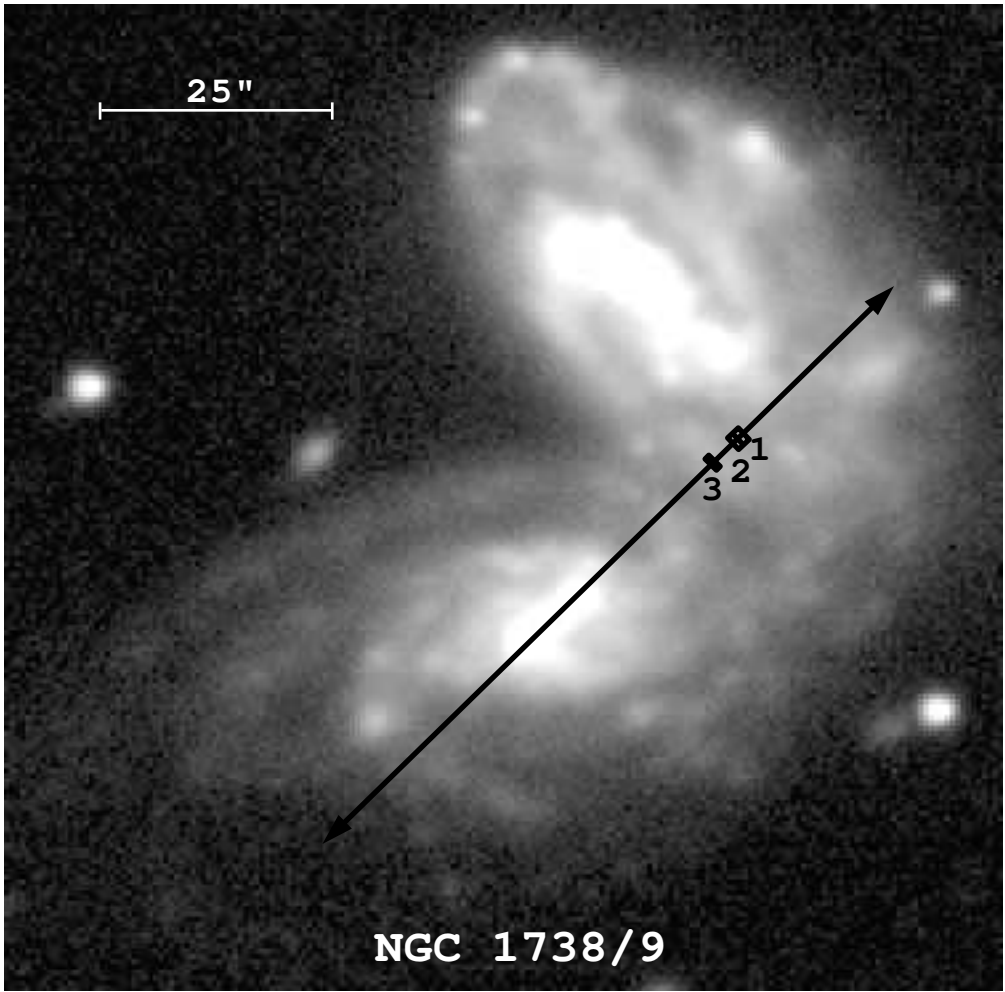
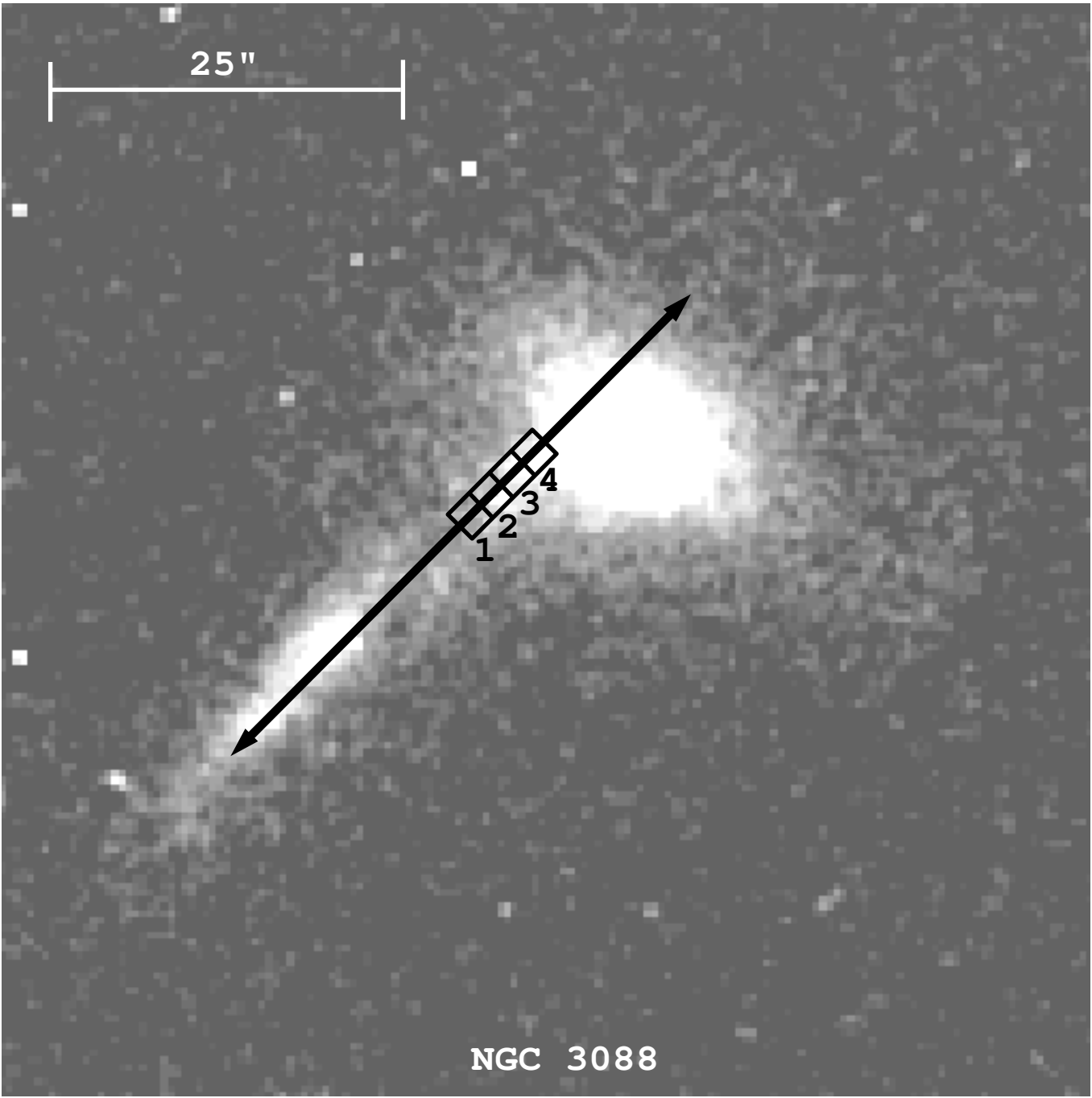
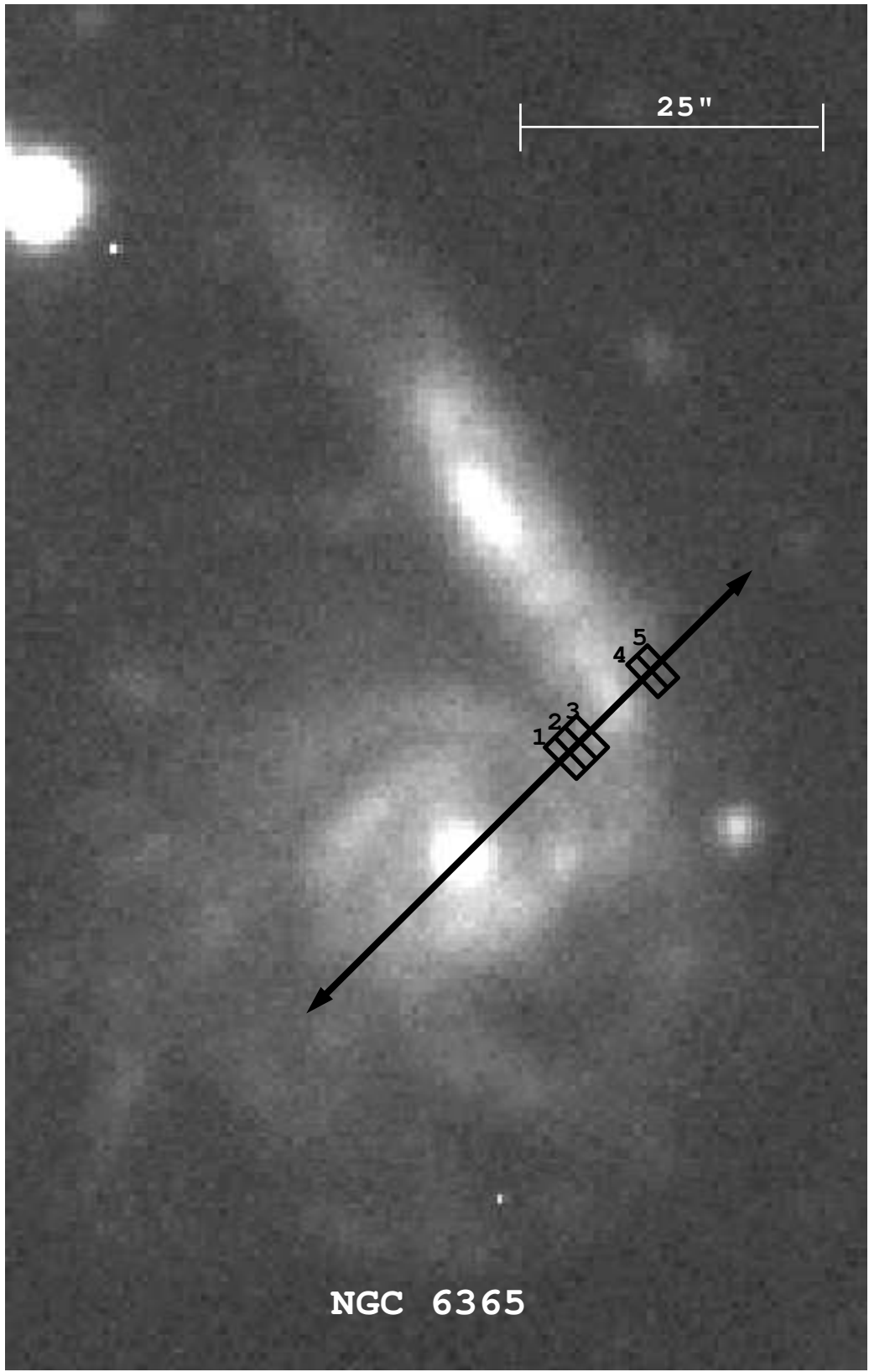
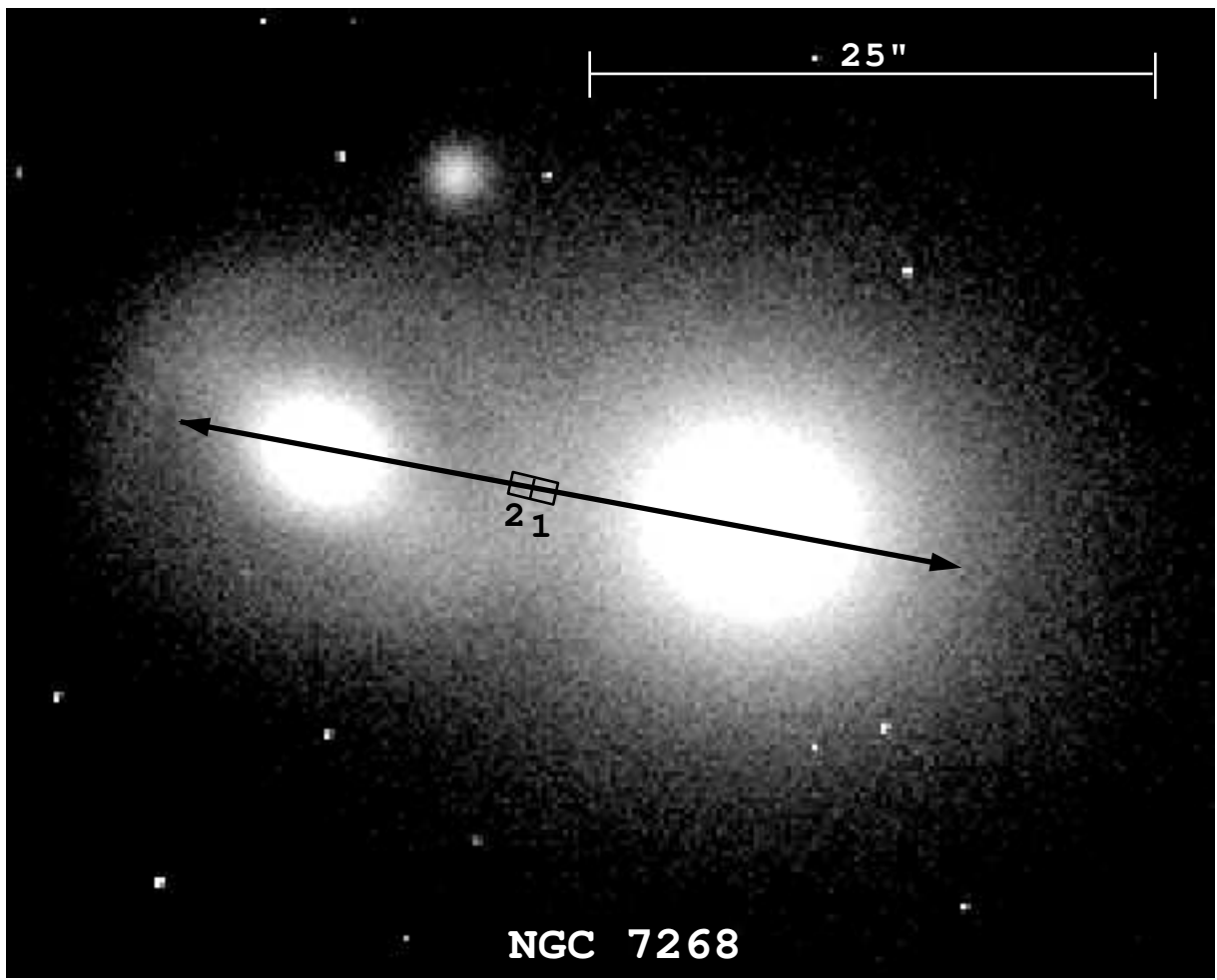
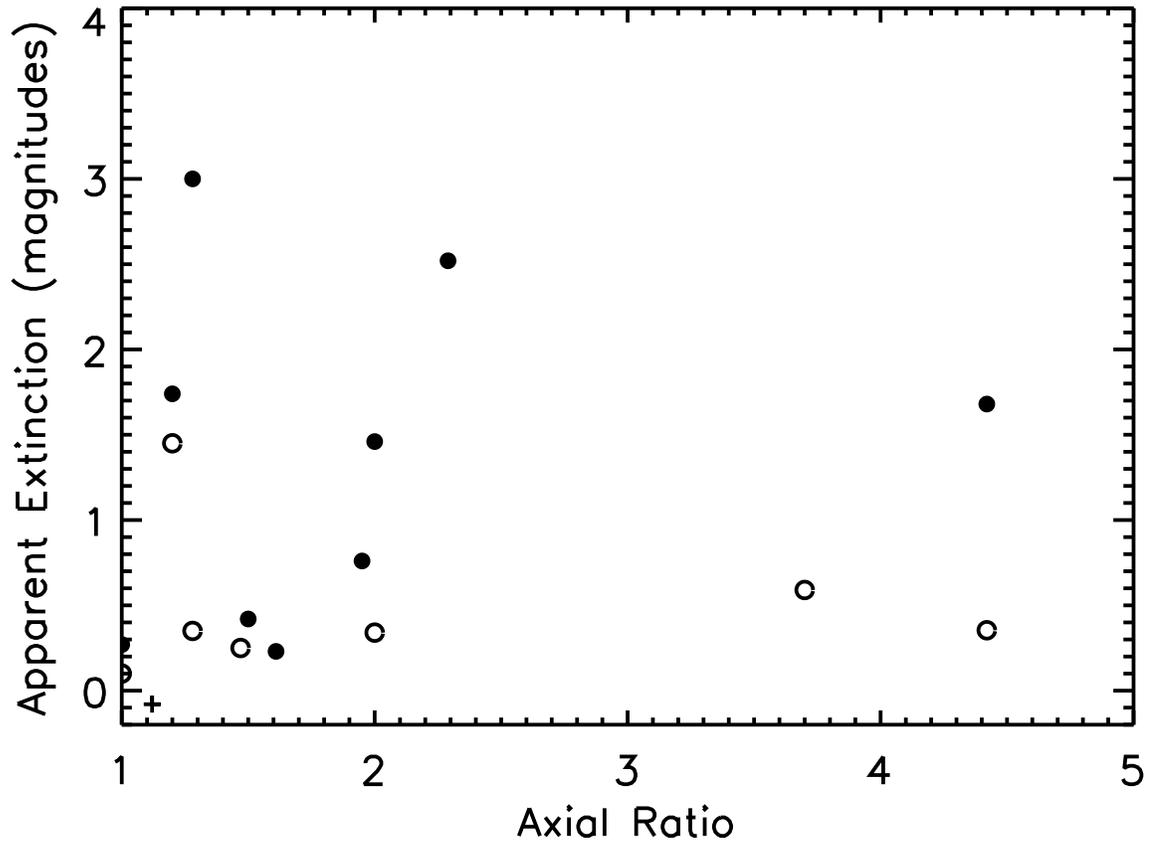
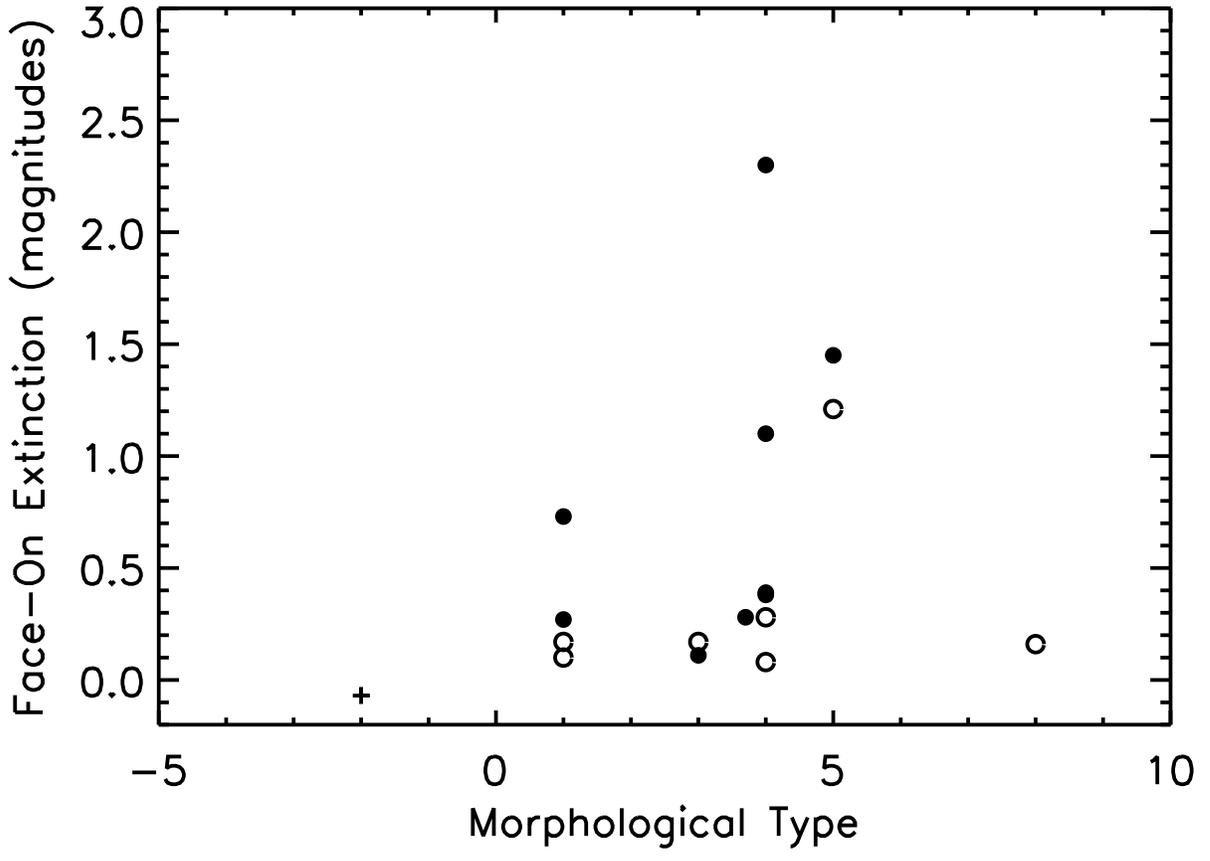
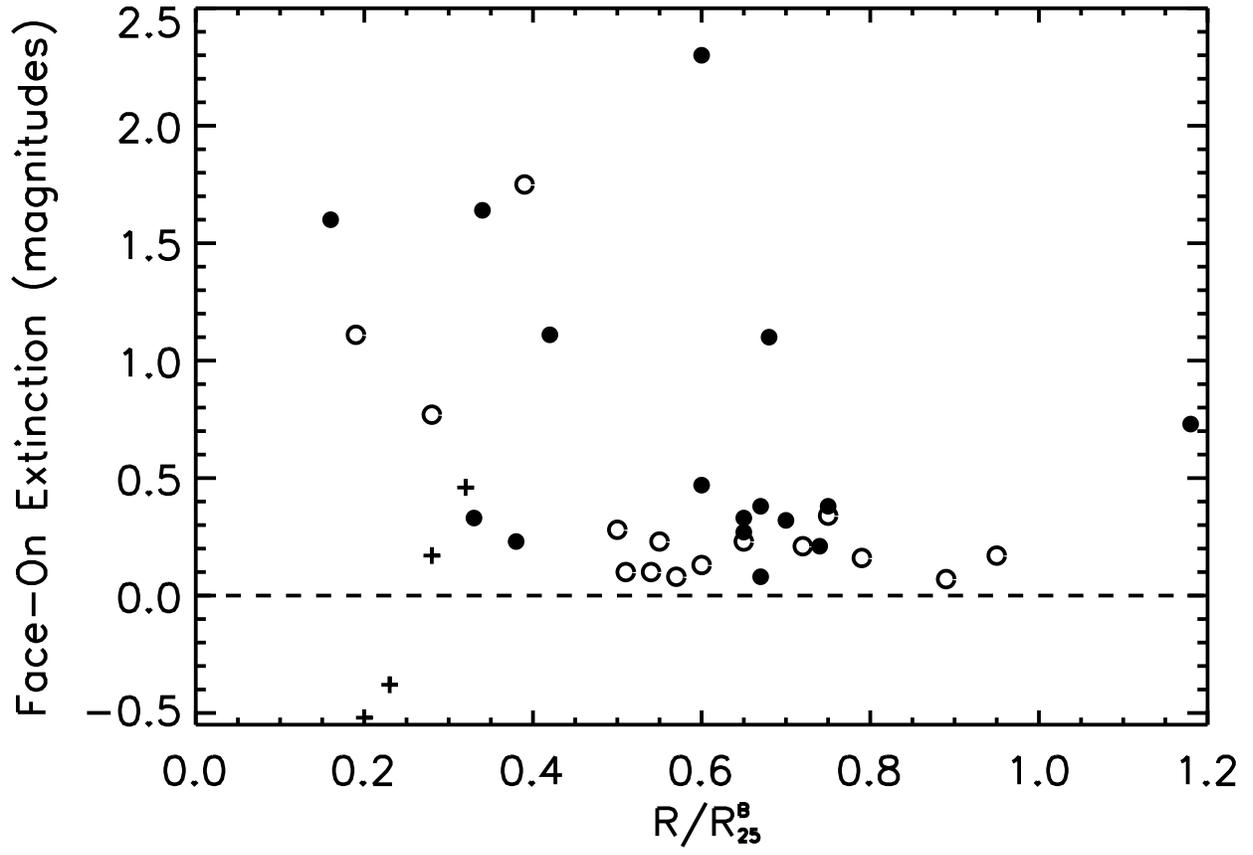
\end{document}